\documentclass{ws-procs975x65}
\pdfoutput=1

\def\chandra    {\emph{Chandra}}
\def\xmm        {\emph{XMM-Newton}}
\def\rosat      {\emph{ROSAT}}
\def\gmrt       {\emph{GMRT}}

\def\kms        {km$\;$s$^{-1}$}
\def\apj        {\emph{ApJ}}
\def\apjl       {\emph{ApJ}}

\def\prd        {\emph{Phys.\ Rev.\ D}}
\def\prl        {\emph{Phys.\ Rev.\ Lett.}}
\def\aap        {\emph{A\&A}}
\def\mnras      {\emph{MNRAS}}
\def\nat        {\emph{Nature}}
\def\figsubcap#1{\par\noindent\centering\footnotesize(#1)}
\def\1e{\mbox{1E\,0657--56}}

\begin{document}

\title{INTERGALACTIC SHOCK FRONTS}

\author{MAXIM MARKEVITCH}

\address{Harvard-Smithsonian Center for Astrophysics,\\
60 Garden St., Cambridge, MA 02138, USA\\
Review talk at 12th Marcel Grossman Meeting, Paris, July 2009 (updated
with 2010 results)} 

\begin{abstract}
When galaxy clusters collide, they generate shock fronts in the hot
intracluster medium.  Observations of these shocks can provide valuable
information on the merger dynamics and physical conditions in the cluster
plasma, and even help constrain the nature of dark matter. To study shock
fronts, one needs an X-ray telescope with high angular resolution (such as
\chandra), and be lucky to see the merger from the right angle and at the
right moment.  As of this writing, only a handful of merger shock fronts
have been discovered and confirmed using both X-ray imaging and gas
temperature data --- those in \1e, A520, A754, and two fronts in A2146. A
few more are probable shocks awaiting temperature profile confirmation ---
those in A521, RXJ\,1314--25, A3667, A2744, and Coma.  The highest Mach
number is 3 in \1e, while the rest has $M \simeq 1.6-2$.  Interestingly, all
these relatively weak X-ray shocks coincide with sharp edges in their host
cluster's synchrotron radio halos (except in A3667, where it coincides with
the distinct radio relic, and A2146, which does not have radio data yet).
This is contrary to the common wisdom that weak shocks are inefficient
particle accelerators, and may shed light on the mechanisms of relativistic
electron production in astrophysical plasmas.
\end{abstract}


\bodymatter

\section{Introduction}
\label{sec:intro}

\begin{figure}%
\begin{center}
 \parbox{2.1in}{%
   \includegraphics[width=5.2cm,bb=63 93 359 389,clip]{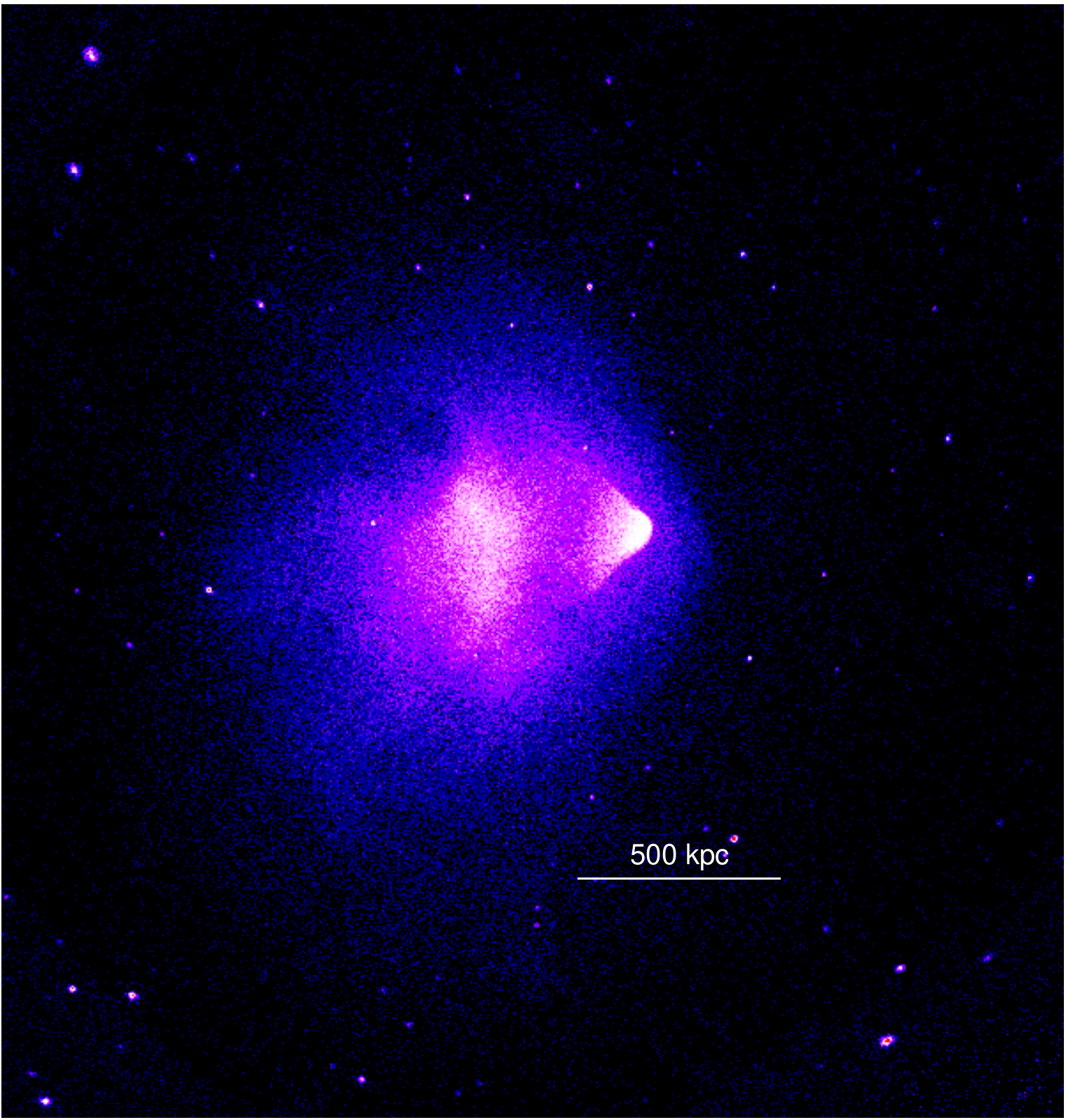}
 \figsubcap{a}
 }
 \hspace*{4pt}
 \parbox{2.1in}{%
   \includegraphics[width=5.2cm]{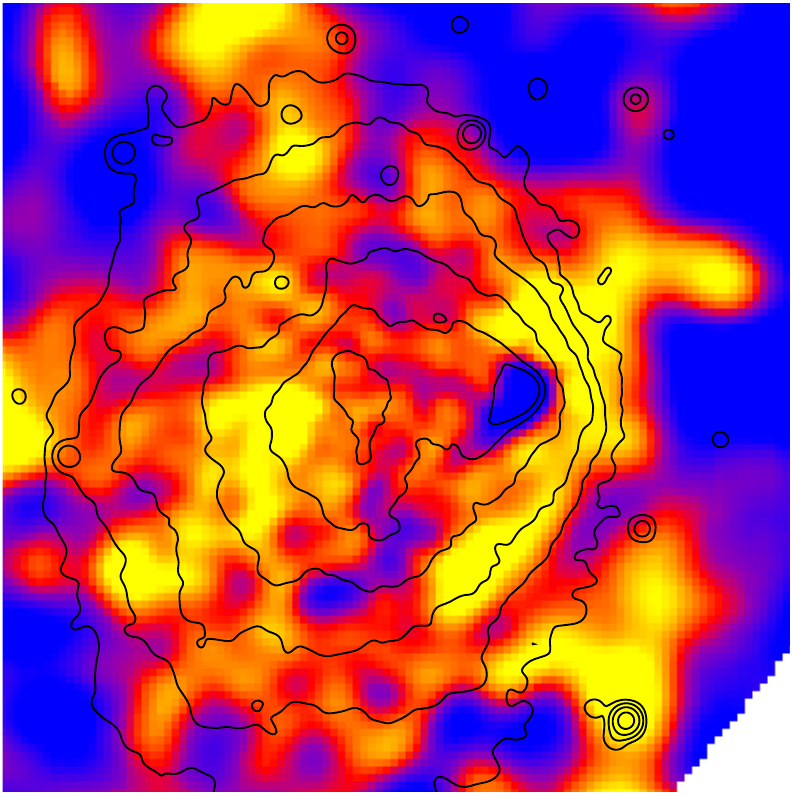}
 \figsubcap{b}
 }
 \caption{The Bullet cluster \1e. (a) \chandra\ X-ray image (0.8--4
   keV band, $7'$ size), (b) gas temperature map (color; blue is $T<6$ keV
   and yellow is $T>20$ keV) and X-ray brightness (contours). A cool gas
   bullet is preceded by a hot shock front.}
\label{fig:1e_img}
\end{center}
\end{figure}

\begin{figure}%
\begin{center}
 \parbox{2.1in}{%
   \includegraphics[width=5.2cm]{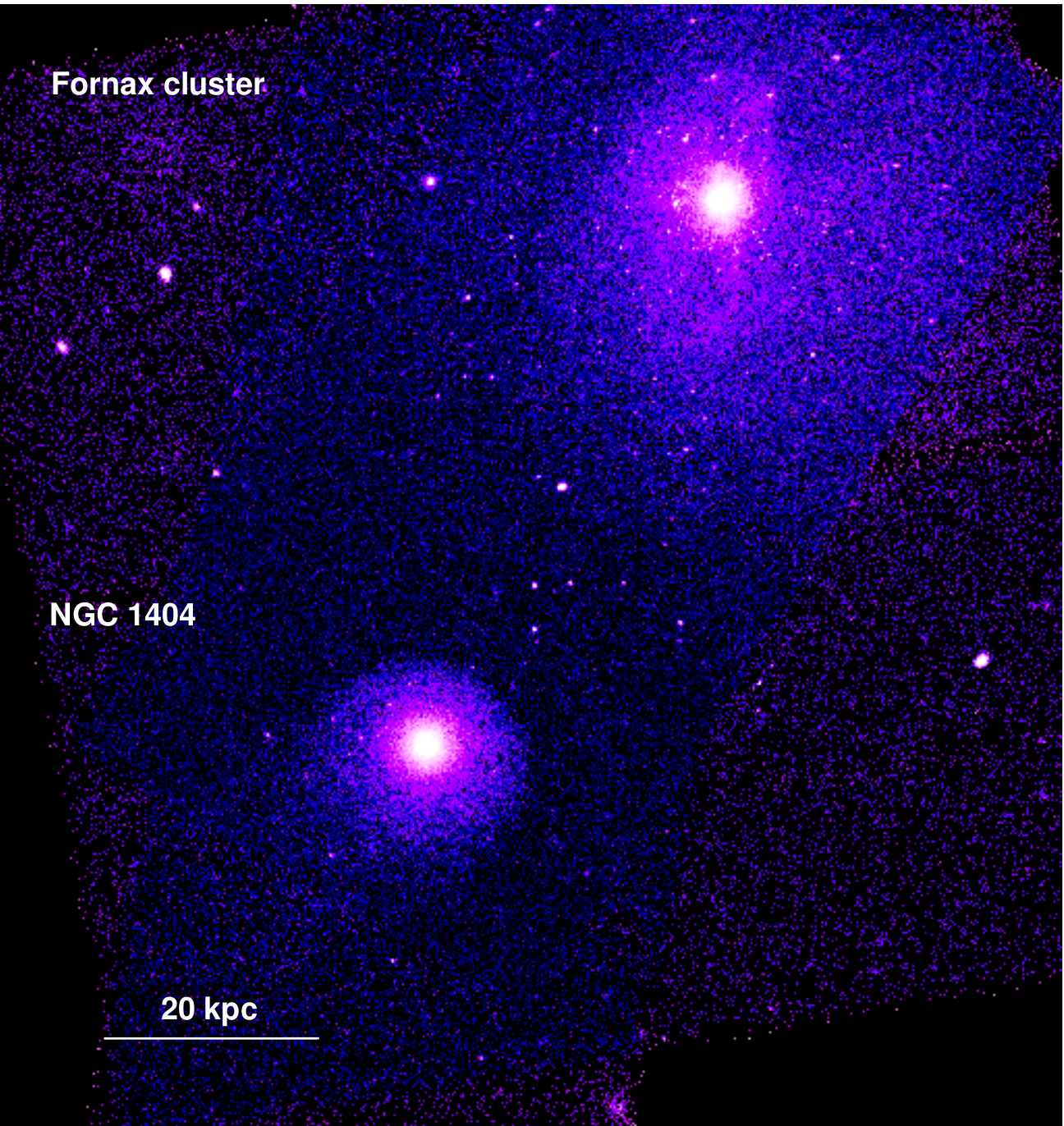}
 \figsubcap{a}
 }
 \hspace*{4pt}
 \parbox{2.1in}{%
   \includegraphics[height=5.2cm]{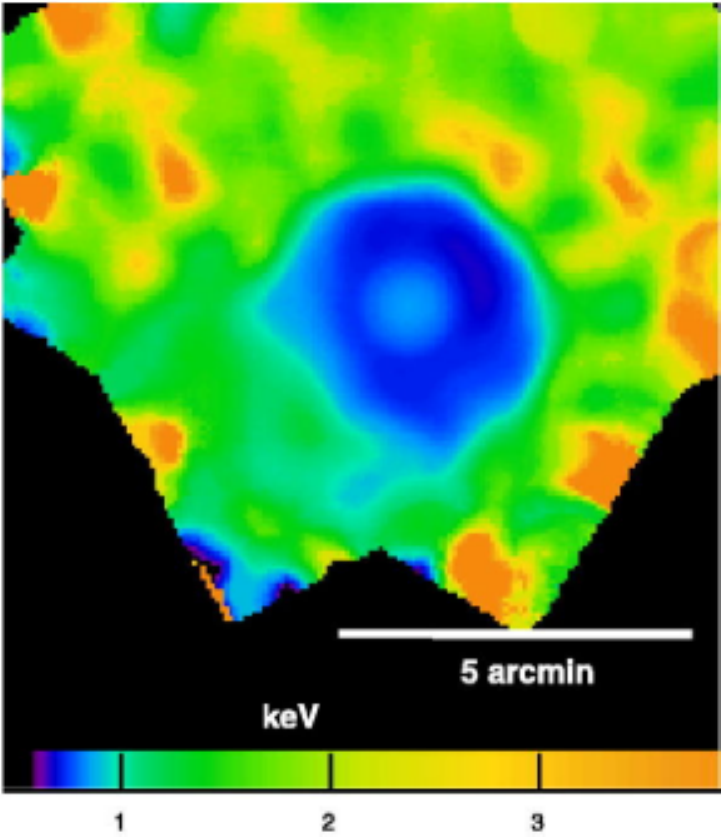}
 \figsubcap{b}
 }
 \caption{Elliptical galaxy NGC\,1404 falling toward the center of  the Fornax
   cluster. (a) \chandra\ X-ray image (0.5--2 keV band, $17'$ size), (b) gas
   temperature map\cite{machacek05} (a zoom-in on NGC\,1404). The sharp NW
   boundary of the galaxy is a cold front.}
\label{fig:n1404}
\end{center}
\end{figure}

Mergers of galaxy clusters provide a unique laboratory to study the cluster
physics --- they generate shocks, hydrodynamic instabilities and turbulence
in the hot intracluster medium (ICM). They also cause ``cold fronts'', or
contact discontinuities, around the cores of infalling subclusters or in the
disturbed cool core of the main cluster\cite{mv07}. In turn, these ICM
disturbances distort and amplify magnetic fields (which are frozen into the
ICM) and generate ultrarelativistic electrons that produce diffuse
synchrotron emission of cluster-wide radio
halos\cite{cassano09,sarazin04,brunetti09}. Of these phenomena, with the
current instruments, we can directly observe shock fronts and cold fronts as
sharp brightness and temperature edges in high angular resolution X-ray
images --- Fig.\ \ref{fig:1e_img} shows a poster child of the merger shocks,
the ``bullet cluster'' \1e\cite{mm02,mm06,mm10}, and Fig.\ \ref{fig:n1404}
shows a cold front in the ``cannonball galaxy'' NGC\,1404\cite{machacek05}.
In the radio, the synchrotron emission is a product of density of
ultrarelativistic electrons and $B^2$, where the magnetic field $B$\/ is
deduced by indirect means\cite{carillitaylor}.  Disentangling the
relativistic electron density and the magnetic field (e.g., by observing
inverse Compton emission from the radio halos), as well as direct
observation of turbulence in the ICM, are beyond the current observing
capabilities.

\begin{figure}
\begin{center}
\includegraphics[width=3.0in,bb=10 30 328 398,clip]{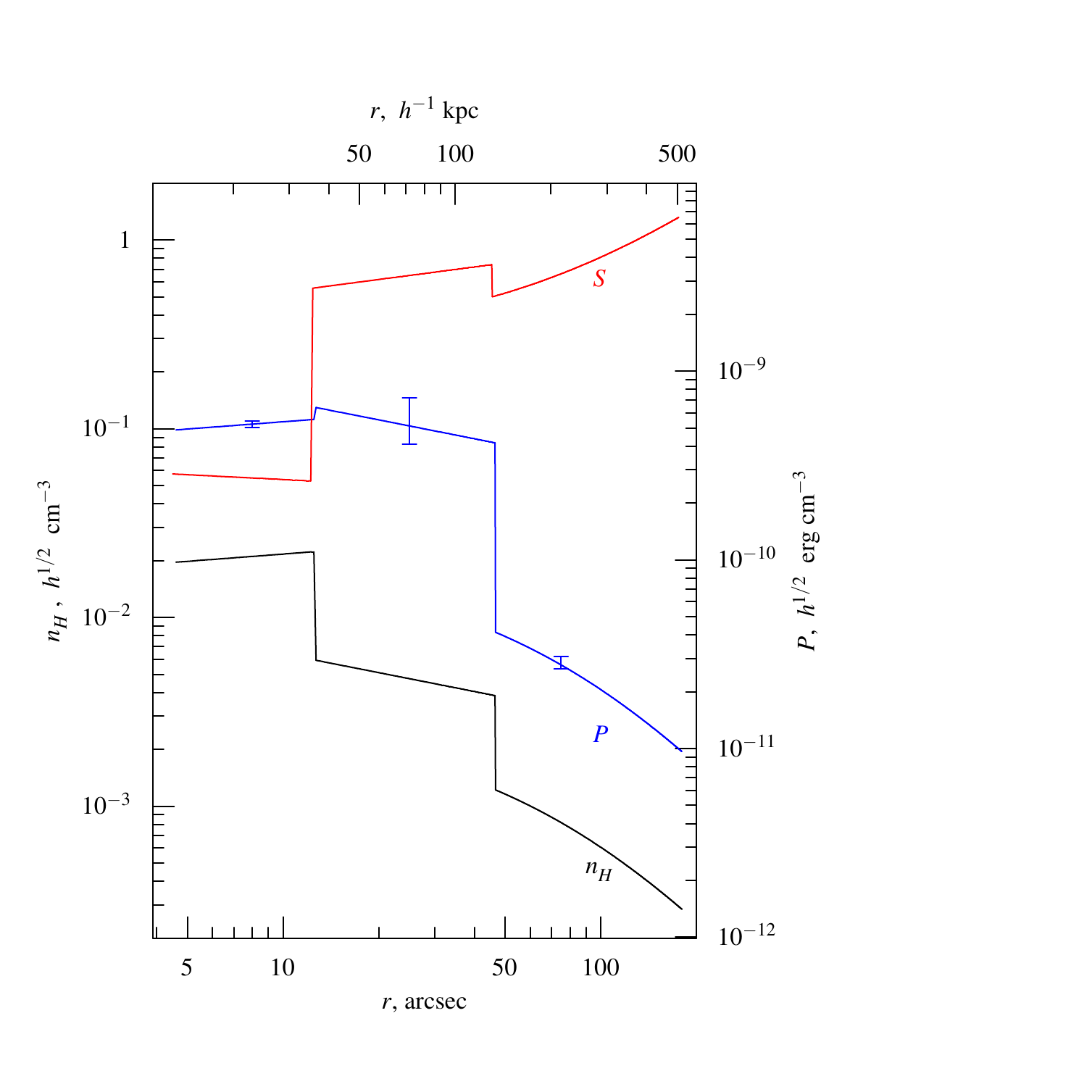}
\end{center}
\caption{Schematic gas density, pressure and specific entropy profiles in a
  sector of the Bullet cluster crossing the bullet nose and the shock front.
  The density jump around $r=12''$ is a cold front (the front boundary of
  the gas bullet), and the jump at $r=50''$ is a shock front.}
\label{fig:1e_profs}
\end{figure}

Figure \ref{fig:1e_profs} schematically shows the ICM density, pressure and
specific entropy profiles in a sector in front of the ``bullet'' in the
Bullet cluster. The sector crosses a cold front (the boundary of the bullet)
and the shock that propagates ahead of it (Fig.\ \ref{fig:1e_img}).  While
both these X-ray brightness edges exhibit similar density jumps, the
pressure across the cold front is almost continuous, as expected for a
contact discontinuity between the dense, low-entropy remnant of an infalling
subcluster and the surrounding shock-heated gas. The shock front exhibits a
large pressure jump and a modest entropy increase, corresponding to its
relatively low Mach number, $M=3$.

Cold fronts are more easily observable than shock fronts -- they were first
discovered with \chandra\ in a merging cluster A2142\cite{mm00} and
subsequently observed in most mergers.  More surprisingly, they are also
seen in the cool cores of at least half of all relaxed
clusters\cite{mm03,mv07}.  Merger shock fronts quickly move away from the
central, bright cluster regions into the faint outskirts, where they are
difficult to observe in X-rays.  To be discernible in an X-ray image, they
also require a merger to occur almost exactly in the plane of the sky. For
these reasons, only a handful is currently known, and they are the subject
of this review.  There are other types of shock fronts expected in clusters
--- far in the outskirts, a strong shock should separate the virialized
cluster region and the outside infalling matter, but the X-ray brightness at
those distances is far too low for these shocks to be observable at present.
Explosions of the central AGNs also generate shocks in the cluster cores,
usually with $M\approx 1$, but sometimes much stronger\cite{croston09}.
These shocks are outside the scope of this review.

\section{Constraints on dark matter from the Bullet cluster}

\begin{figure}
\begin{center}
\includegraphics[width=5.2cm]{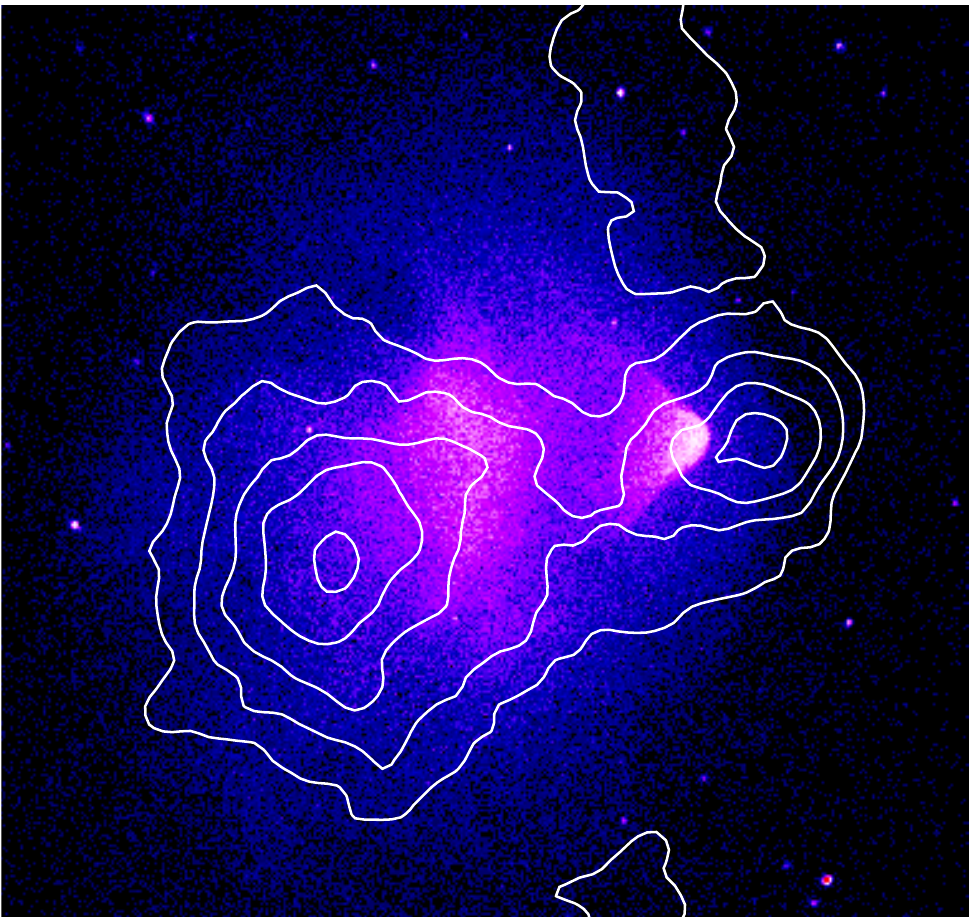}
\end{center}
\caption{X-ray image of \1e (same as in Fig.\ \ref{fig:1e_img}a) with
  contours showing total projected mass from weak lensing. The gas
  subcluster trails its mass peak because of ram pressure. This offset is
  direct proof of the existence of dark matter\cite{clowe06}.}
\label{fig:1e_lens}
\end{figure}

We will start with some physical results already derived using the most
prominent cluster shock front, the one in the Bullet cluster \1e.  Fig.\
\ref{fig:1e_img} shows an X-ray image and projected temperature map derived
from a 500 ks \chandra\ exposure\cite{mm10,million09,owers09} of \1e, and
Fig.\ \ref{fig:1e_lens} overlays a weak lensing total mass map\cite{clowe06}
on the \chandra\ image.  It is clear from these figures that \1e\ is a
merger of two subclusters flying apart just after a collision.  The cooler
gas bullet, with its characteristic shuttlecock shape, appears to be a
remnant of the cool core of the smaller subcluster, pushed back from its
host dark matter peak by ram pressure. The X-ray brightness and temperature
profiles across the shock front give a Mach number ($M=3.0\pm 0.4$) and a
velocity of the shock front (4700 \kms)\cite{mm06}.  The velocity of the
subcluster is probably lower than that of the shock front, because the
pre-shock gas flows in from the cluster outskirts toward the shock front
under the gravitational pull of the subcluster\cite{springel07}. The two
subclusters should have passed through each other just 200 million years
ago.

It has been quickly realized\cite{clowe04,clowe06} that the offset between
the peaks of the total mass and the gas mass peaks seen in Fig.\
\ref{fig:1e_lens} offers the first direct and model-independent evidence for
the existence of dark matter, as opposed to some forms of modified gravity,
where the visible matter is all there is and the laws of gravity are
incorrect on large scales\cite{milgrom83,bekenstein04} --- an alternative
possibility put forward to explain the longstanding problem of ``missing
mass'' in galaxy clusters\cite{zwicky37}.  Indeed, the X-ray emitting gas is
by far the dominant visible mass component in clusters, and yet, as the
lensing mass map shows, the peaks of the {\em total}\/ mass density are
clearly located elsewhere. The measured total masses in those peaks are as
expected from the ratio of gas mass, stellar mass in the galaxies, and total
mass normally observed in clusters. What makes this cluster unique is the
moment at which it is caught by the observer --- when gas and dark matter
have been temporarily spatially separated by a violent merger, revealing
that these are indeed two different kinds of matter.

The survival of two dark matter subclusters after a near-direct collision
(indicated by the X-ray data) also provides an upper limit on
self-interaction cross-section for the dark matter particles. The dark
matter peaks coincide with peaks of the galaxy number density, and both dark
matter and galaxy peaks are located ahead of the gas, as expected for
collisionless dark matter and galaxies vs.\ fluid-like ICM. The observed
mass-to-light ratio within the gas-depleted subcluster peaks is found to be
similar to that in other clusters within similar radii.  This excludes the
possibility that a significant fraction of dark matter particles has escaped
the subclusters as a result of particle collisions during the subclusters'
recent passage through each other. Taking into account the merger velocity
derived from the X-ray, this observation places a limit of $\sigma/m<0.7$
cm$^2$/g on the dark matter self-interaction
cross-section.\cite{mm04,randall08}, excluding most of the astrophysically
interesting range of the cross-section.\cite{spergel00}.

\section{Electron-proton equilibration timescale in the ICM}

\begin{figure}[b]%
\begin{center}
 \parbox{2.1in}{%
  \includegraphics[width=5.2cm]{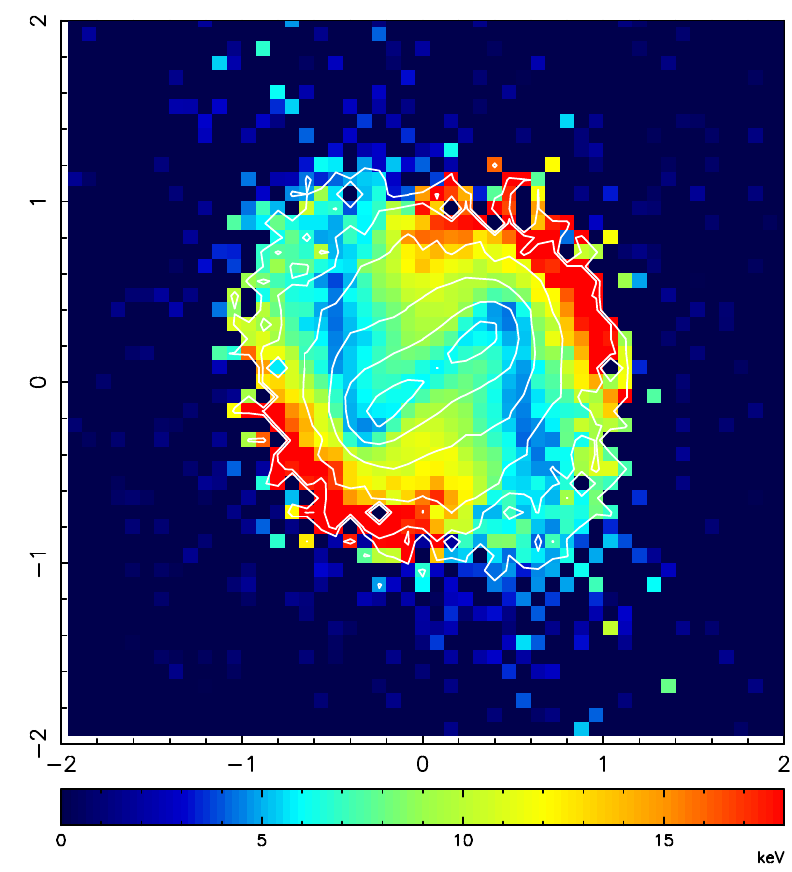}
 \figsubcap{a}
 }
 \hspace*{4pt}
 \parbox{2.1in}{%
  \includegraphics[width=5.2cm]{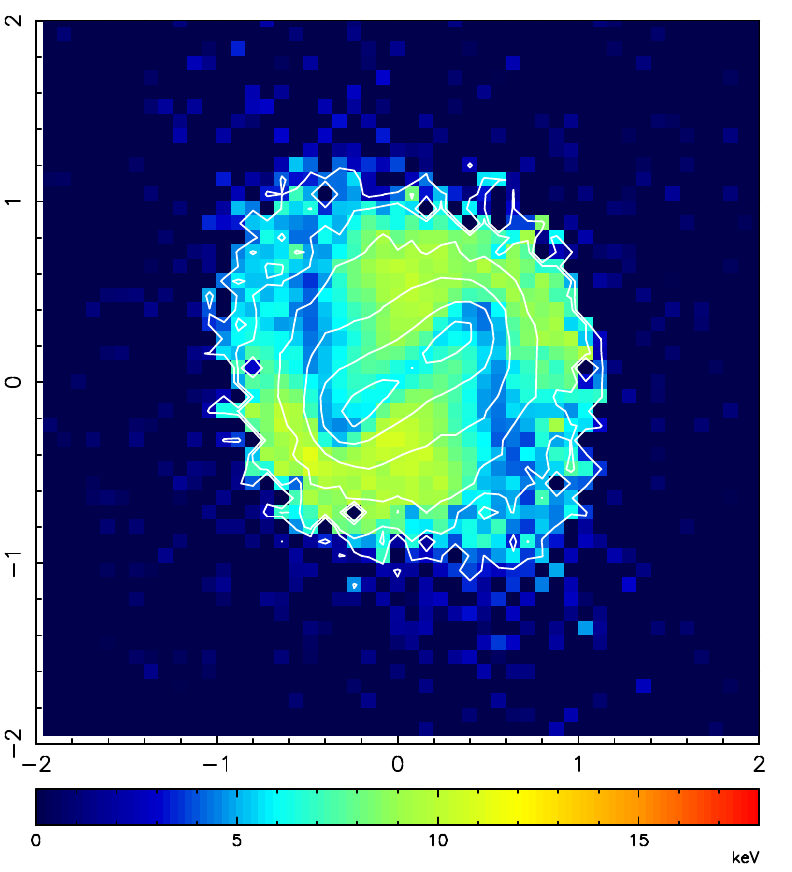}
 \figsubcap{b}
 }
 \caption{Simulations of a cluster merger\cite{takizawa99}, with colors
   showing the temperature and contours showing the mass. There are two
   shock fronts propagating ahead of two subclusters that have just crossed
   each other. (a) Average plasma temperature, (b) electron temperature
   assuming Coulomb electron-proton equilibration timescale.  The linear
   scale is in Mpc and the temperature scale is in keV. The X-ray measured
   $T_e$ may be a significant underestimate of thermodynamic temperature at
   shocks.}
\label{fig:takiz}
\end{center}
\end{figure}

\begin{figure}[t]
\begin{center}
\includegraphics[height=2.5in,bb=27 43 532 542,clip]{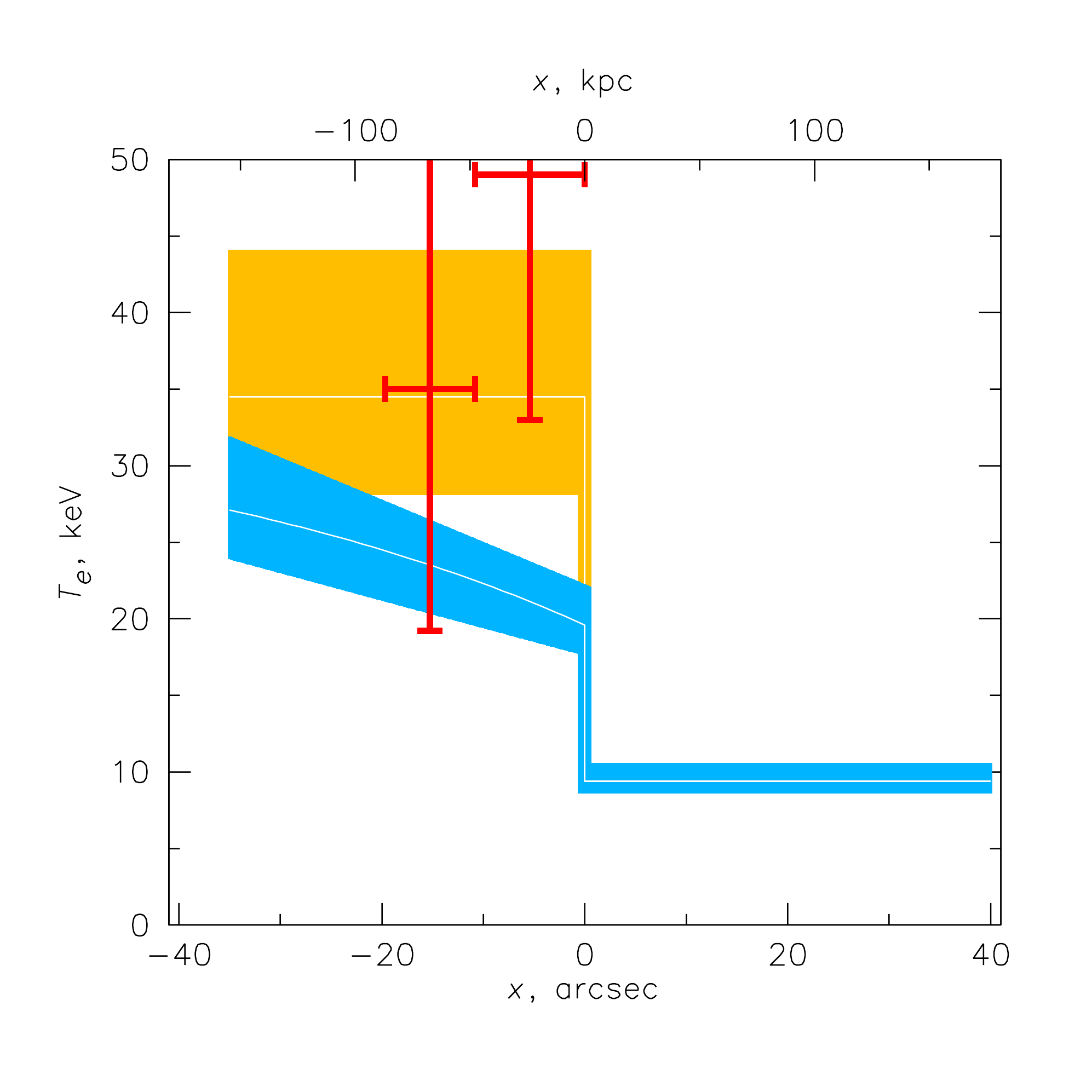}
\end{center}
\caption{Predicted $T_e$ profiles for the shock in \1e, for the Coulomb
  electron-proton equilibration timescale (blue band) and for instant
  equilibration (yellow band).  Overlaid is the \chandra\ measurement 
  (deprojected, with $1\sigma$ error bars), which indicate a
  shorter equilibration timescale\cite{mm06}.}
\label{fig:1e_tei}
\end{figure}

The passage of a merger shock in a fully ionized intracluster plasma should
heat the protons dissipatively, while electrons (for shocks with $M\ll
(m_p/m_e)^{1/2}\approx 43$, which is always true for cluster mergers) are
compressed adiabatically and subsequently heated by other mechanisms, such
as Coulomb collisions with hotter protons.  The linear sizes and
temperatures of galaxy clusters are such that on a timescale of collisional
electron-proton equilibration, $\tau_{ep}$, the shock travels significant
distances, creating observable regions where the electron temperature $T_e$
(the only one that we can currently measure in X-rays) is below the true
thermodynamic temperature.  This is illustrated in Fig.\ \ref{fig:takiz},
which shows simulated maps of the average and electron temperatures for a
merging cluster with two shocks propagating outwards, assuming collisional
$\tau_{ep}$. The $T_e/T_p$ nonequilibrium is expected in may types of
astrophysical objects, from solar wind to supernovae to WHIM filaments.
However, it is usually impossible to measure $T_e$ and $T_p$ independently
{\em and}\/ map them on linear scales on which their equilibrium is expected
to be achieved.  Shock fronts in clusters provide a unique opportunity to do
so and thus constrain $\tau_{ep}$.  Because cluster shocks are relatively
weak, the easily measurable plasma density jump at the shock is sufficiently
far from its asymptotic value (factor of 4 for a $\gamma=5/3$), which allows
one to derive the Mach number of the shock from the density jump using the
Rankine-Hugoniot jump conditions, and predict the equilibrium post-shock
value of the plasma temperature independently of the measured jump in $T_e$.
The pre-shock $T_e$\/ gives the sound speed, which, together with the shock
density jump, gives the speed of the post-shock gas flow relative to the
shock front. This can be used to predict the post-shock $T_e$\/ profile for
various values of $\tau_{ep}$.  For the shock in \1e\ with $M=3$, the
expected rise in post-shock $T_e$\/ for the case of Coulomb collisions can
be spatially resolved by \chandra, as shown in Fig.\ \ref{fig:1e_tei}. The
measured values of the post-shock $T_e$ exclude the Coulomb timescale at a
95\% significance, favoring a much shorter $\tau_{ep}$.\cite{mm06}

\section{Other clusters with known shock fronts}

\begin{figure}[t]%
\begin{center}
 \parbox{2.1in}{%
  \includegraphics[width=5.2cm]{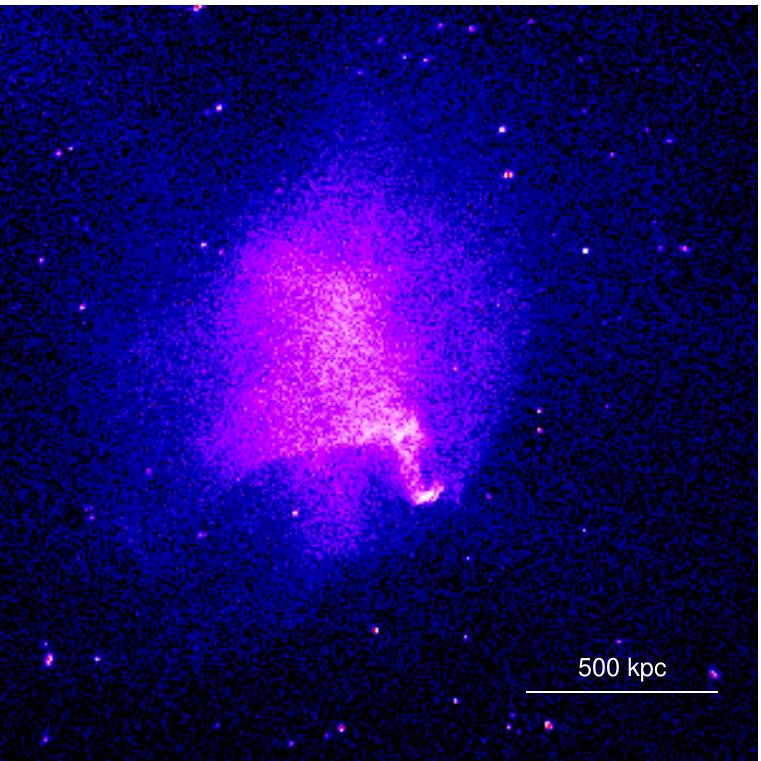}
 \figsubcap{a}
 }
 \hspace*{4pt}
 \parbox{2.1in}{%
   \includegraphics[width=5.2cm,bb=120 130 340 350,clip]{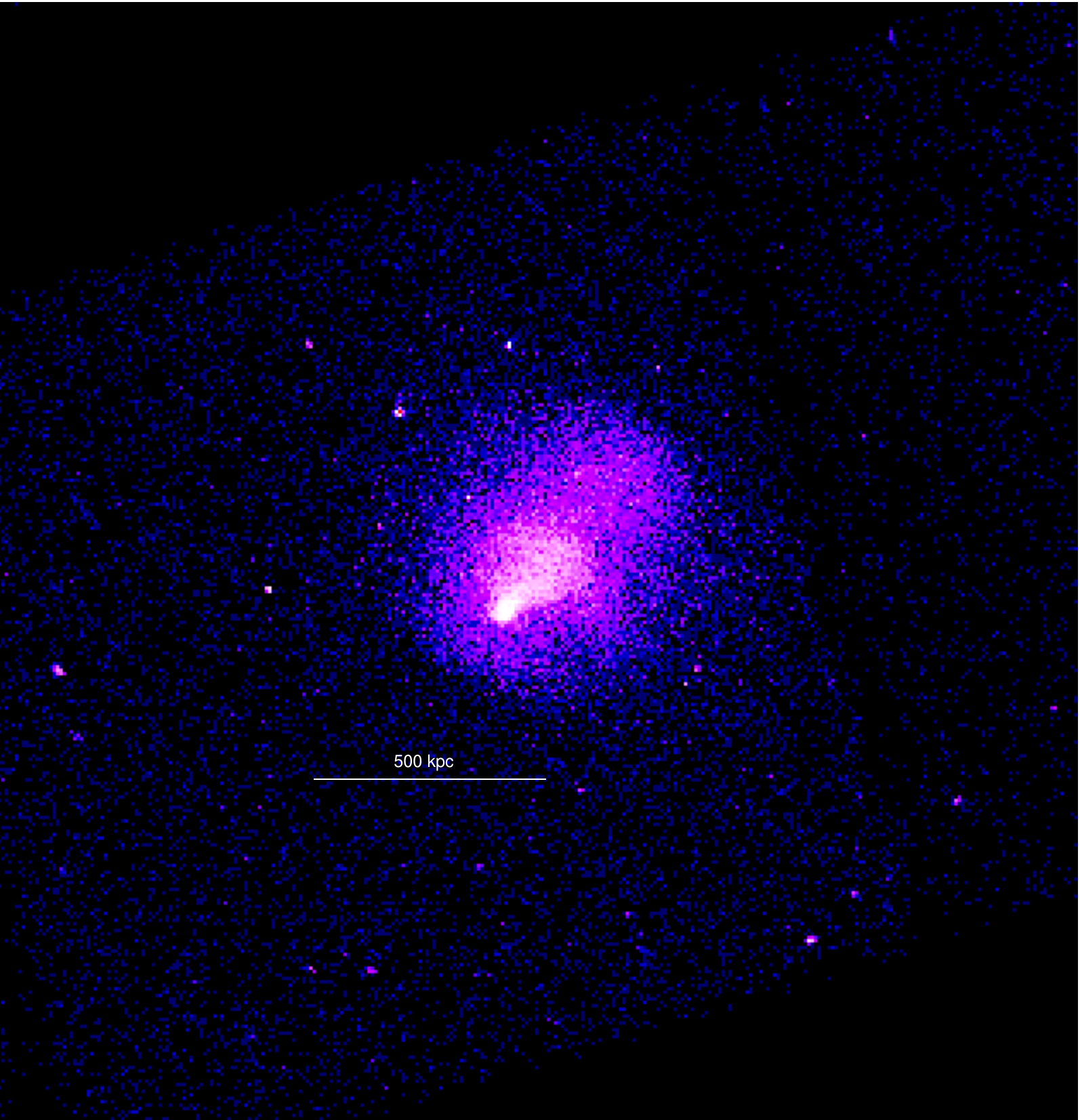}
 \figsubcap{b}
 }
 \caption{(a) \chandra\ image of A520; the shock front with $M=2$ is to the
   SW of the bright central structure\cite{mm05}. (b) \chandra\ image of
   A2146, showing two $M\simeq 2$ shocks about 300 kpc SE and NW of the
   center.\cite{russell10}} 
\label{fig:a520_a2146}
\end{center}
\end{figure}

Until recently, only two merger shock fronts were known and confirmed by the
X-ray temperature measurement --- the \1e\ discussed above and A520 shown in
Fig.\ \ref{fig:a520_a2146}a. The latter has a shock with $M\simeq
2$\cite{mm05}; its recent long \chandra\ observation is currently being
analyzed to validate the above constraint on $\tau_{ep}$.  A520 also
exhibits a similar offset between gas and dark matter\cite{clowe10}.

Recently, several more shock fronts and shock candidates were found in
X-rays. A unique case of two fronts located on the opposite sides of the
cluster has been reported in A2146\cite{russell10}, shown in Fig.\
\ref{fig:a520_a2146}b --- a geometry expected for a symmetric merger with a
small impact parameter. Both shocks have $M\approx 2$. Similar second shocks
are not observed in either \1e\ or A520, most likely because those shocks
have already moved out of the bright central regions in these mergers of
subclusters of very different masses.

\begin{figure}[h]%
\begin{center}
 \parbox{2.1in}{%
  \includegraphics[height=5.2cm,bb=24 14 277 267,clip]{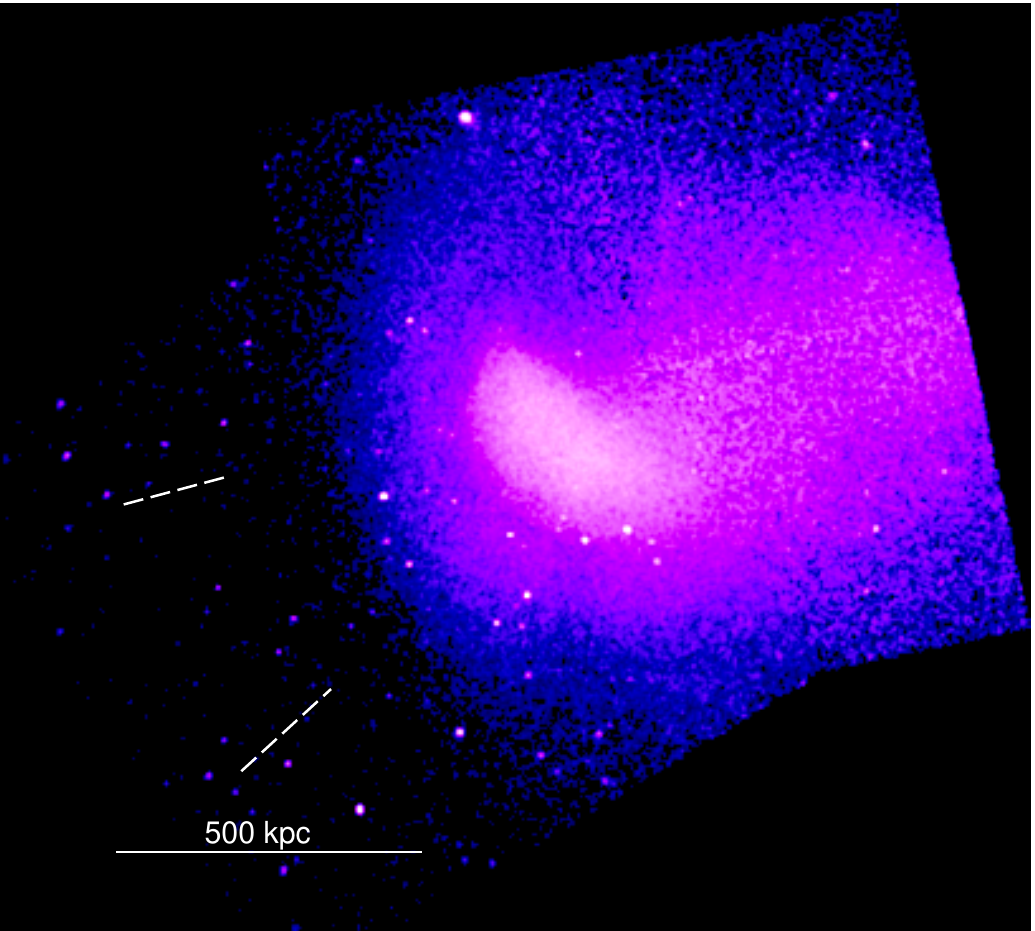}
 \figsubcap{a}
 }
 \hspace*{4pt}
 \parbox{2.1in}{%
   \includegraphics[height=5.2cm,bb=16 41 530 573,clip]%
    {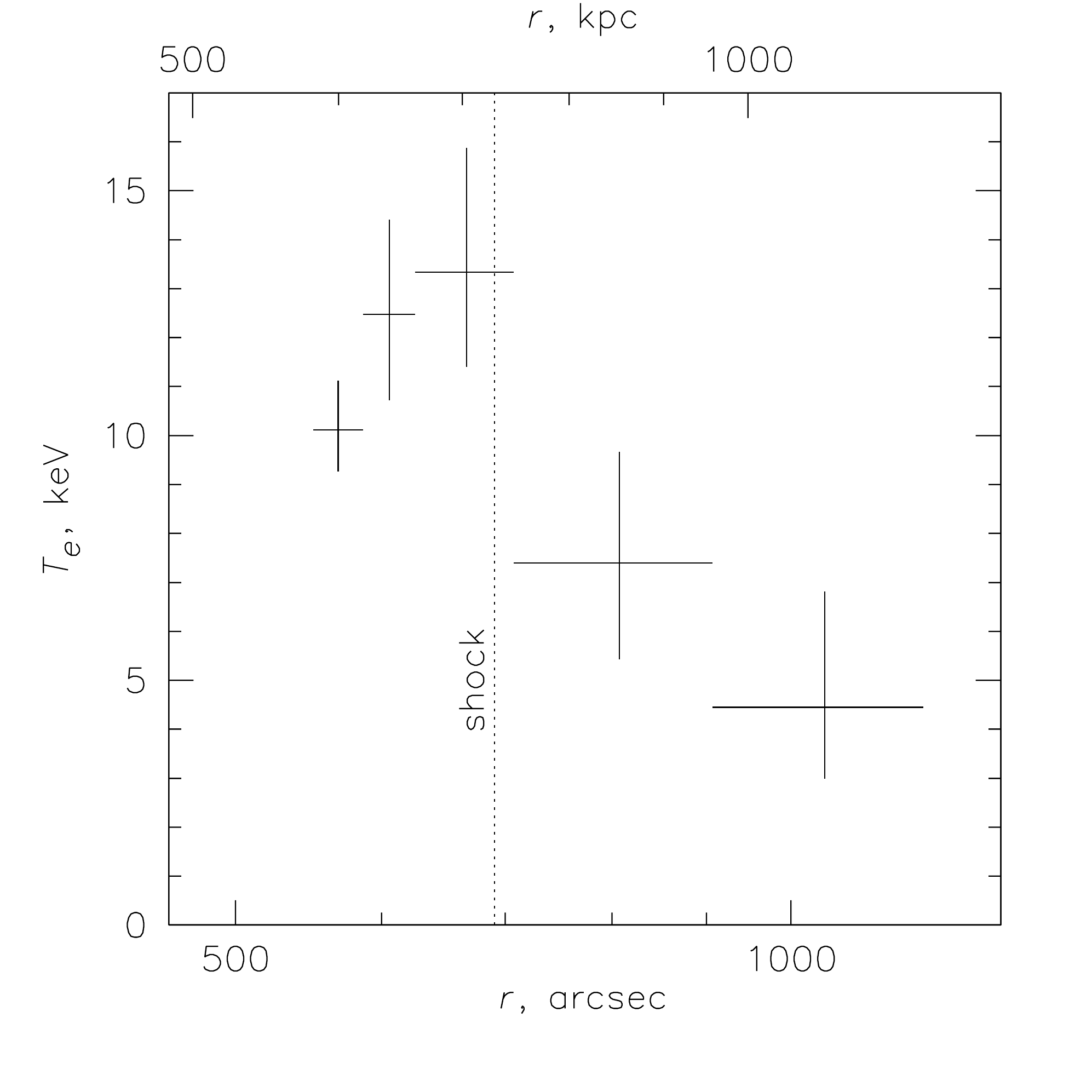}
 \figsubcap{b}
 }
 \caption{(a) \chandra\ image of A754 with the sector showing a shock
   front with $M\simeq 1.6$\cite{krivonos03,macario10}. (b) \chandra\ gas
   temperature profile showing a temperature jump across the front in a
   sector shown in panel (a) (radius is from the shock's center of
   curvature), confirming that this is indeed a shock.\cite{macario10}}
\label{fig:a754}
\end{center}
\end{figure}

Now that we know how the merger shocks look like, it is possible to find
them in archival X-ray data of lower quality. A possible shock front was
detected in the \rosat\ PSPC image of A754\cite{krivonos03} and recently
confirmed using the \chandra\ temperature measurement across the
shock\cite{macario10}, as shown in Fig.\ \ref{fig:a754}.  This is a
relatively weak shock with $M=1.6$.

\begin{figure}%
\begin{center}
 \parbox{2.1in}{%
  \includegraphics[height=5.2cm]{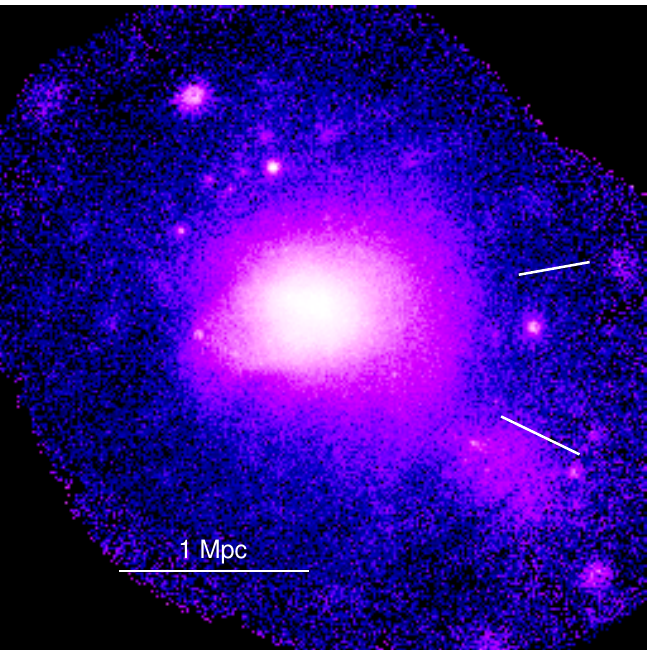}
 \figsubcap{a}
 }
 \hspace*{4pt}
 \parbox{2.1in}{%
   \includegraphics[height=5.2cm,bb=13 40 512 500,clip]{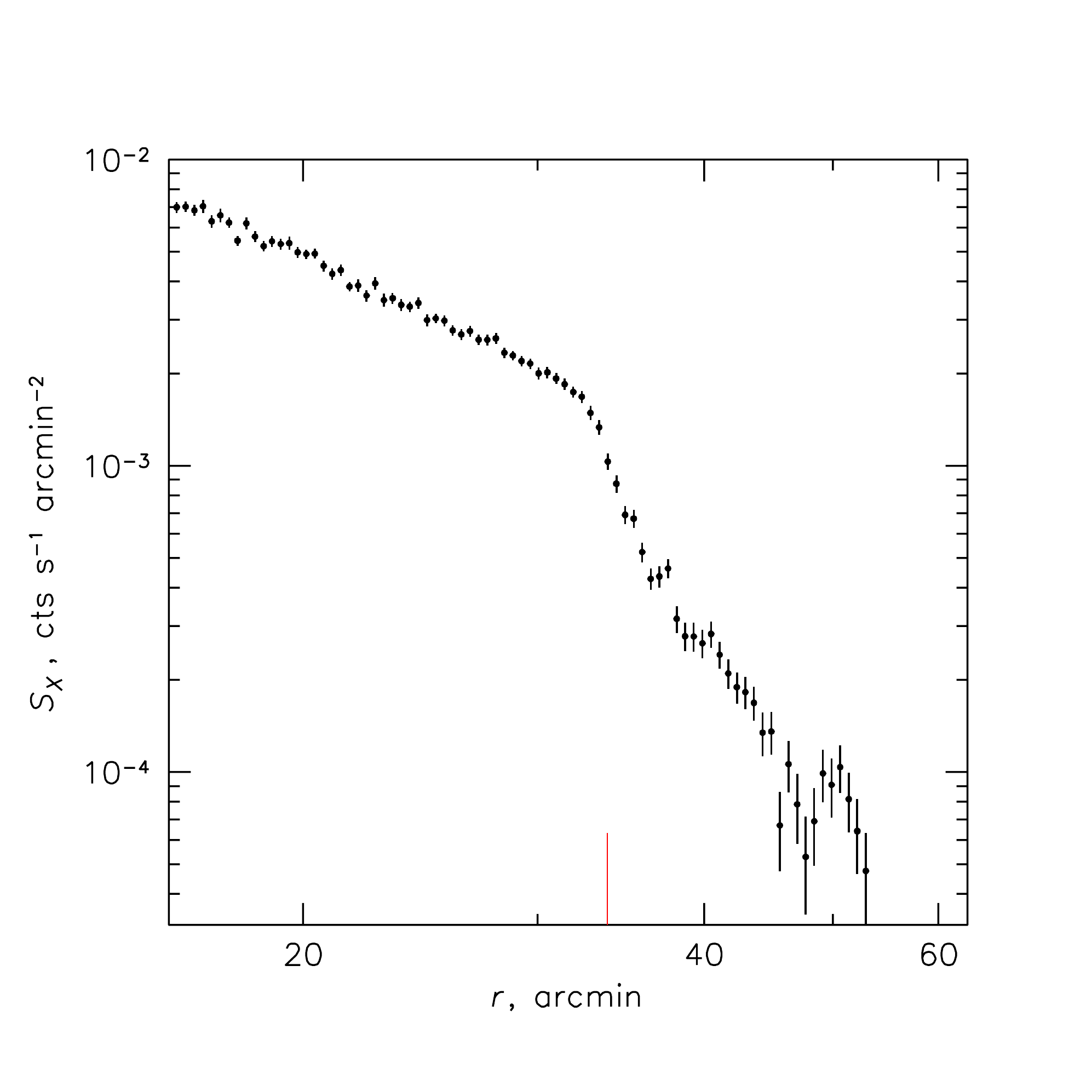}
 \figsubcap{b}
 }
 \caption{(a) \rosat\ PSPC mosaic of Coma ($1.6^\circ\times1.6^\circ$),
   showing a brightness edge in the eastern sector (shown by dashes) that
   might be a shock front. (b) X-ray brightness profile in that sector,
   showing this feature at around $r=33'$ (red dash).}
\label{fig:coma}
\end{center}
\end{figure}

\begin{figure}%
\begin{center}
 \parbox{2.1in}{%
  \includegraphics[height=5.2cm]{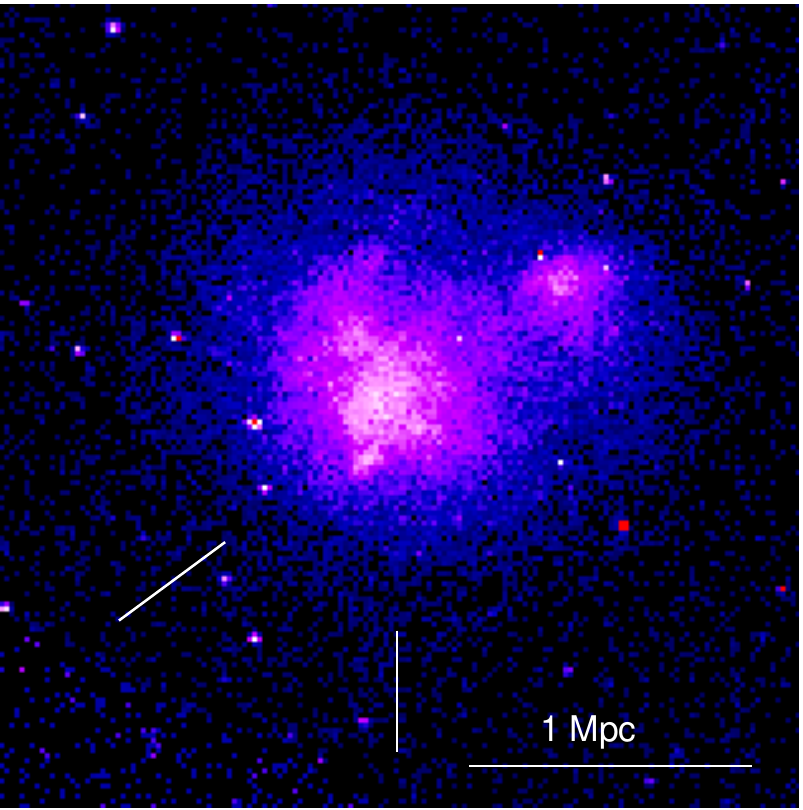}
 \figsubcap{a}
 }
 \hspace*{4pt}
 \parbox{2.1in}{%
   \includegraphics[height=5.2cm,bb=11 43 510 494,clip]%
    {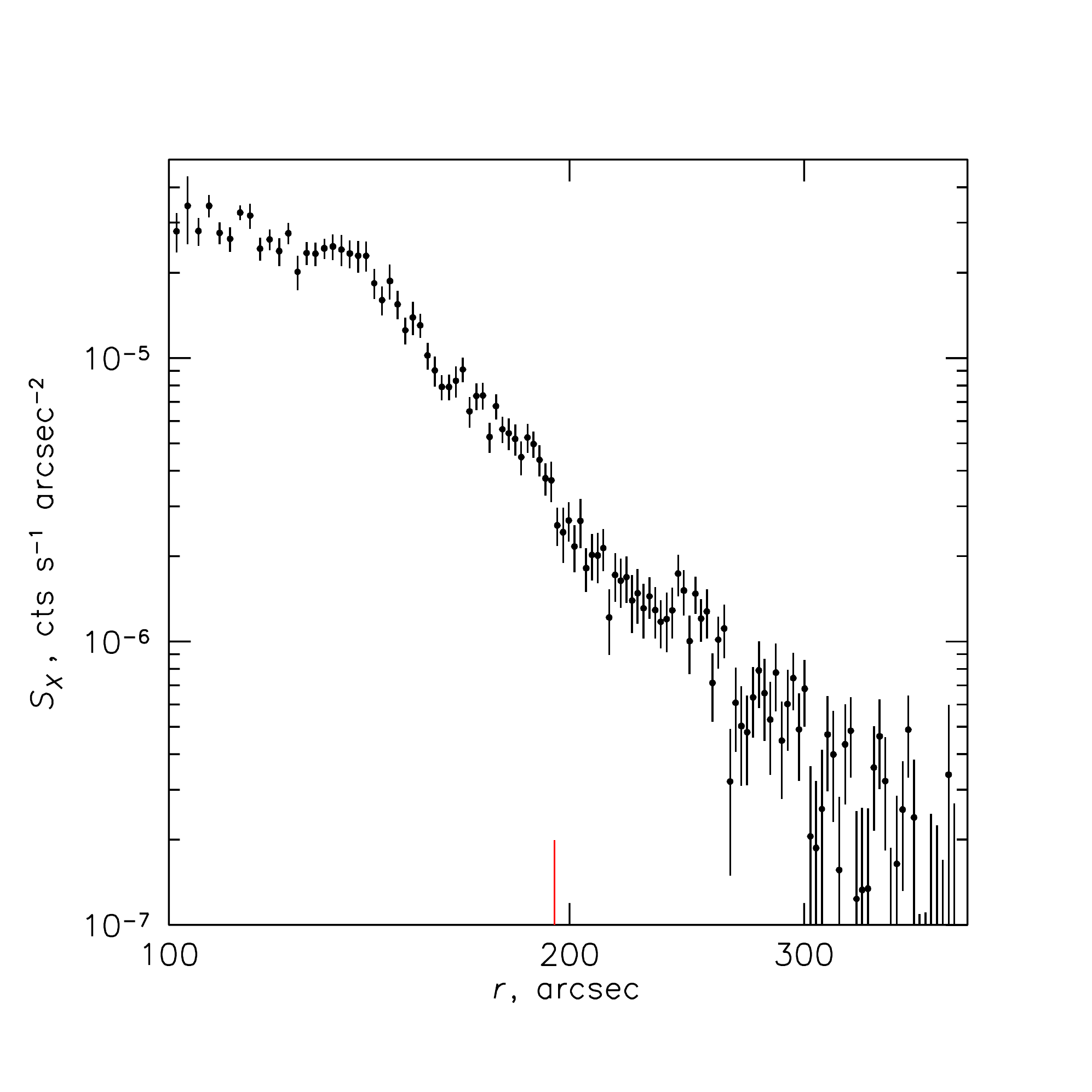}
 \figsubcap{b}
 }
 \caption{(a) \chandra\ image of A2744 ($10'\times 10'$), showing a subtle
   brightness edge in the SE sector (shown by dashes) that might be a shock
   front. (b) X-ray brightness profile in that sector, showing this feature
   at around $r=190''$ (red dash).}
\label{fig:a2744}
\end{center}
\end{figure}

\begin{figure}%
\begin{center}
 \parbox{2.1in}{%
  \includegraphics[height=5.2cm,bb=33 12 333 312,clip]{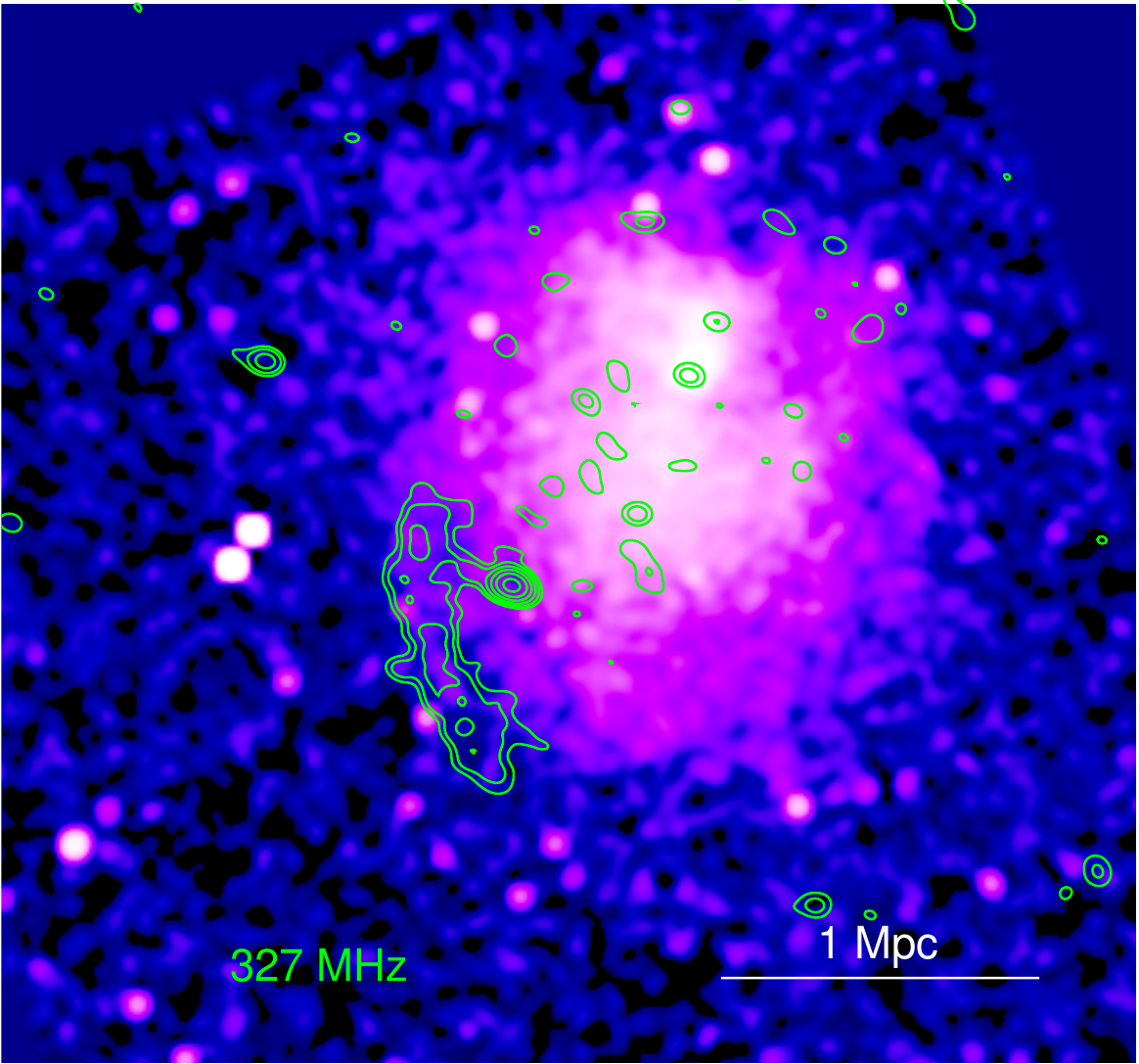}
 \figsubcap{a}
 }
 \hspace*{4pt}
 \parbox{2.1in}{%
   \includegraphics[height=5.2cm,bb=11 42 510 494,clip]%
    {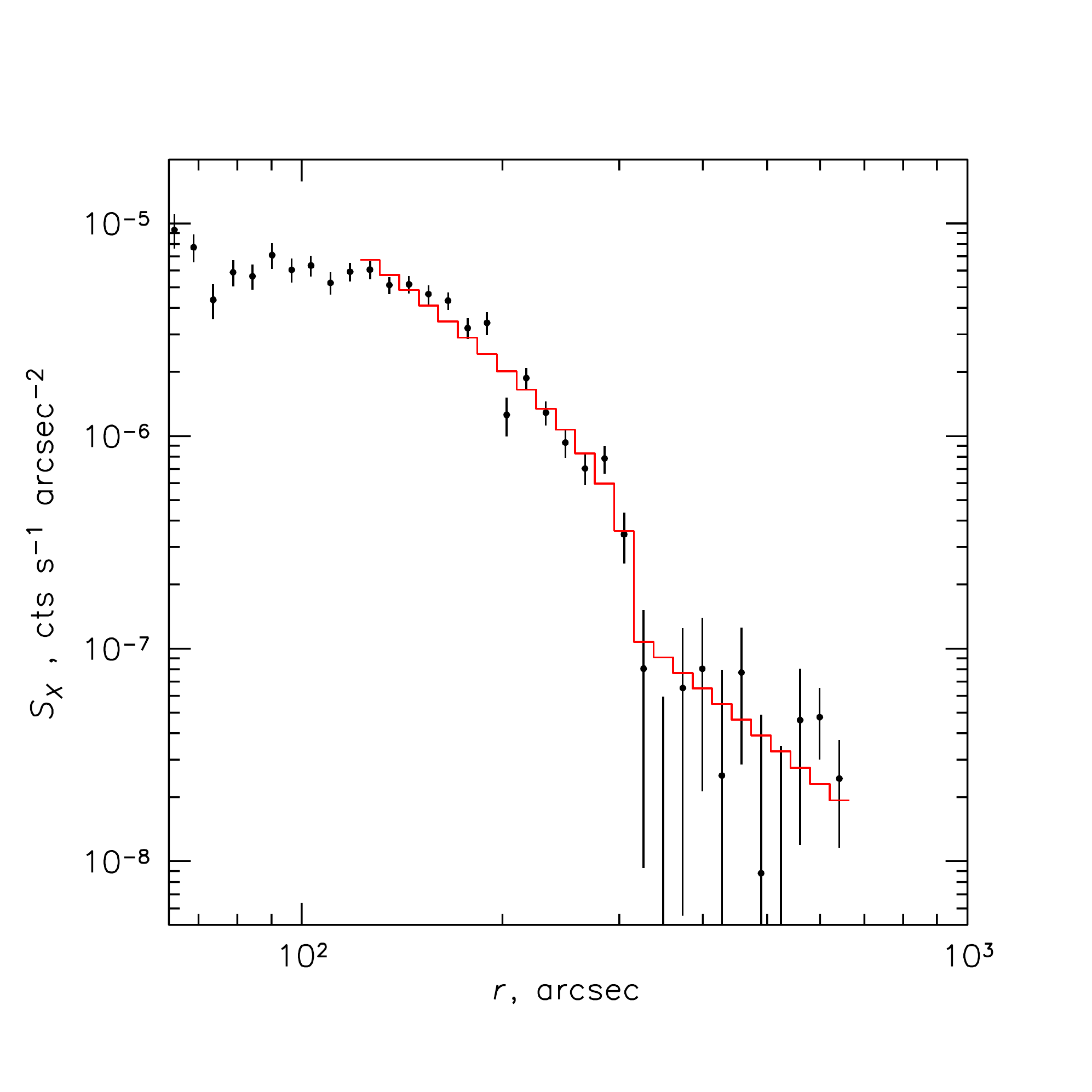}
 \figsubcap{b}
 }
 \caption{(a) \chandra\ image of A521, with \gmrt\ 327 MHz synchrotron
   radio relic contours overlaid\cite{giacintucci08}, delineating a possible
   shock front. (b) \chandra\ X-ray brightness profile in the sector covered
   by the relic; radius is from the relic's center of curvature. Overlaid is
   a model shock profile with $M=2.3$ predicted from the radio spectrum of
   the relic\cite{giacintucci08}, which is consistent with the X-ray imaging
   data. Gas temperature data with sufficient quality is not yet available
   to confirm this shock.}
\label{fig:a521}
\end{center}
\end{figure}

As of this writing (August 2010), these are the only known merger shocks
with the X-ray temperature profiles confirming that an edge-like feature in
the X-ray brightness is indeed a shock front (and not a cold front, for
example).  There are several other {\em likely}\/ shock fronts seen in the
X-ray images but awaiting temperature confirmation.  One is in Coma cluster,
seen in a \rosat\ PSPC mosaic (Fig.\ \ref{fig:coma}).  There is some
indication of a temperature jump of the correct sign in the \xmm\
temperature map\cite{wik09}, though a measurement more matched to the shock
position is needed to verify this.  Coma is big enough for this putative
shock front to be seen by the {\em Planck}\/ observatory that maps the
Sunyaev-Zeldovich signal, which is proportional to gas pressure and thus
should readily reveal shocks.

Another candidate front is found in the \chandra\ image of A2744, a
spectacular merger at $z=0.3$. An X-ray brightness profile in a sector
containing the putative shock is shown in Fig.\ \ref{fig:a2744}. The
putative shock is a very low-contrast feature, likely corresponding to a low
Mach number. We will see below that both Coma and A2744 exhibit edge
features in their radio halo maps that coincide with these putative shocks.

It has long been suggested that peripheral radio relics are caused by
electrons accelerated at shock fronts in the periphery of merging
clusters.\cite{ensslin98}. However, most relics are located far from the
X-ray bright cluster regions, so it is rarely possible to check for the
presence of a shock front in an X-ray image, and even more difficult to get
a confirming temperature measurement.

A likely shock in A521 was discovered by analyzing the \chandra\ X-ray image
at the position of a prominent radio relic\cite{giacintucci08}, see Fig.\
\ref{fig:a521}. There is a subtle X-ray brightness edge right where the
``front'' side of the relic is. The relic has a good radio spectrum, well
represented by a power law. If one assumes that the radio-emitting
relativistic electrons come from Fermi acceleration at the shock, the
spectrum implies $M=2.3$. The density jump for such a shock is in good
agreement with the X-ray brightness profile, as shown by red model in Fig.\
\ref{fig:a521}b. At lower radio frequencies, this cluster reveals a giant
radio halo that starts at the relic and spans the entire
cluster\cite{brunetti09} (Fig.\ \ref{fig:rhalos}, discussed below).

An X-ray brightness edge that looks like a shock was discovered in an \xmm\
observation of the famous NW radio relic in A3667\cite{finoguenov10}; if
this is indeed a shock, the gas density jump indicates $M\approx 2$. A
possible $M\approx 2$ shock front was discovered at the position of a radio
relic in the cluster RXJ\,1314--25\cite{mazzotta10}. The curious M shape of
this front indicates a velocity gradient in the pre-shock gas --- possibly
an extreme case of the effect seen in the hydrodynamic simulations of the
Bullet cluster\cite{springel07}. Table 1 gives a summary of the currently
known merger shocks.

\begin{table}
\tbl{X-ray merger shock fronts and candidates (as of summer 2010)}
{\begin{tabular}{p{1.7cm}ccccl}
\toprule
Cluster     & $\rho$\/ jump & {\em T}\/ jump & {\em M} & Radio edge? & X-ray
refs. \\
\colrule
1E\,0657--56& yes & yes & 3   & yes & \citen{mm02,mm06} \\
A520        & yes & yes & 2   & yes & \citen{mm05} \\
A754        & yes & yes & 1.6 & yes & \citen{krivonos03,henry04,macario10} \\
A2146 N     & yes & yes & 2 & no data & \citen{russell10} \\
A2146 S     & yes & yes & 2 & no data & \citen{russell10} \\
\colrule
A521        & yes &           & 2 & yes & \citen{giacintucci08} \\
RXJ\,1314--25&yes &           & 2 & yes & \citen{mazzotta10} \\
A3667       & yes &           & 2 & yes & \citen{finoguenov10} \\
A2744       & yes &           &   & yes & this work \\
Coma        & yes &           &   & yes & this work \\
\botrule
\end{tabular}}
\label{table:summ}
\end{table}

\section{Radio halos and X-ray shock fronts}

\begin{figure}%
\begin{center}
 \parbox{2.1in}{%
  \includegraphics[height=5.2cm,bb=47 57 266 276,clip]%
    {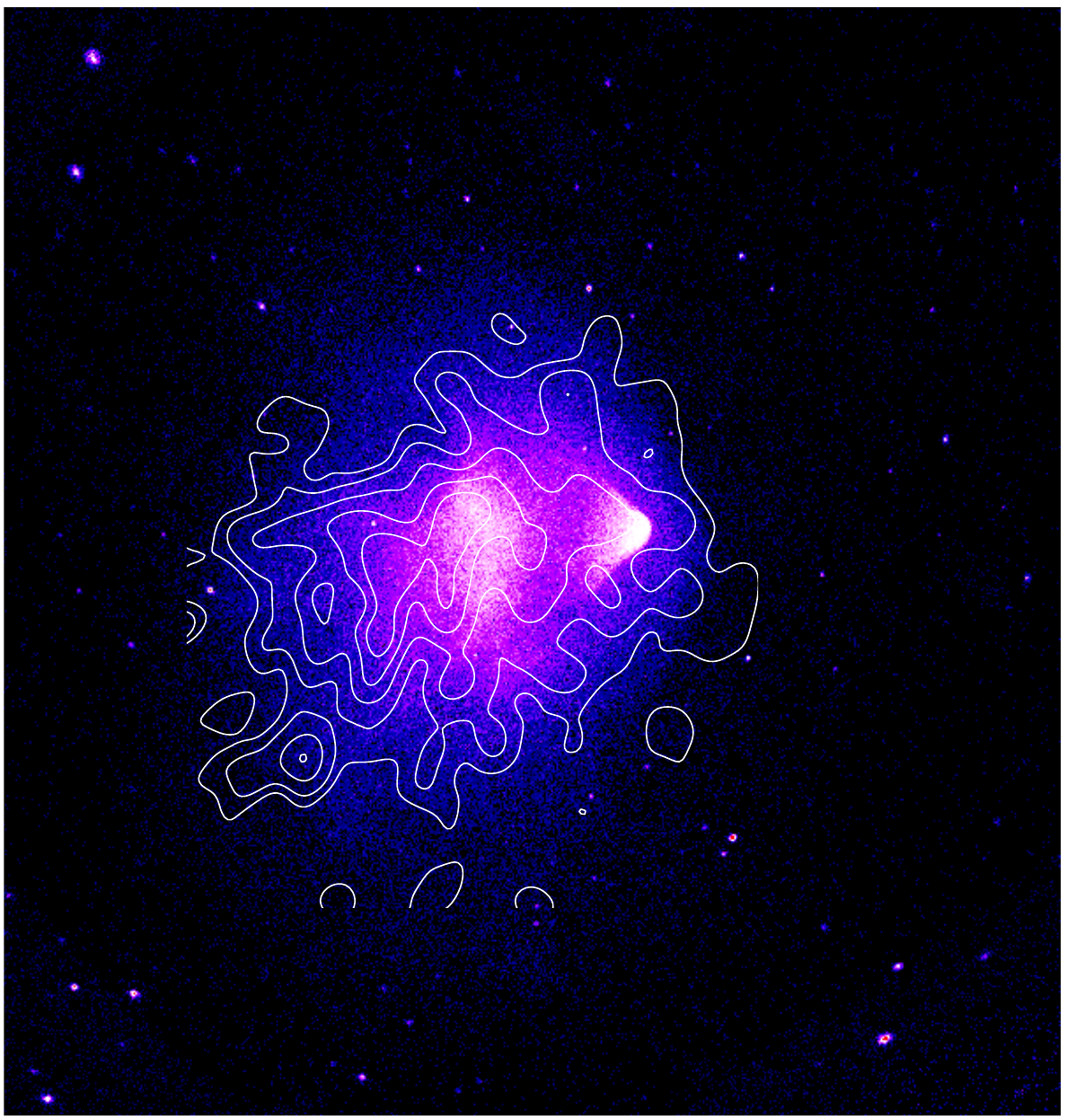}
 \figsubcap{1E\,0657--56}
 }
 \hspace*{4pt}
 \parbox{2.1in}{%
   \includegraphics[height=5.2cm,bb=70 20 395 345,clip]%
     {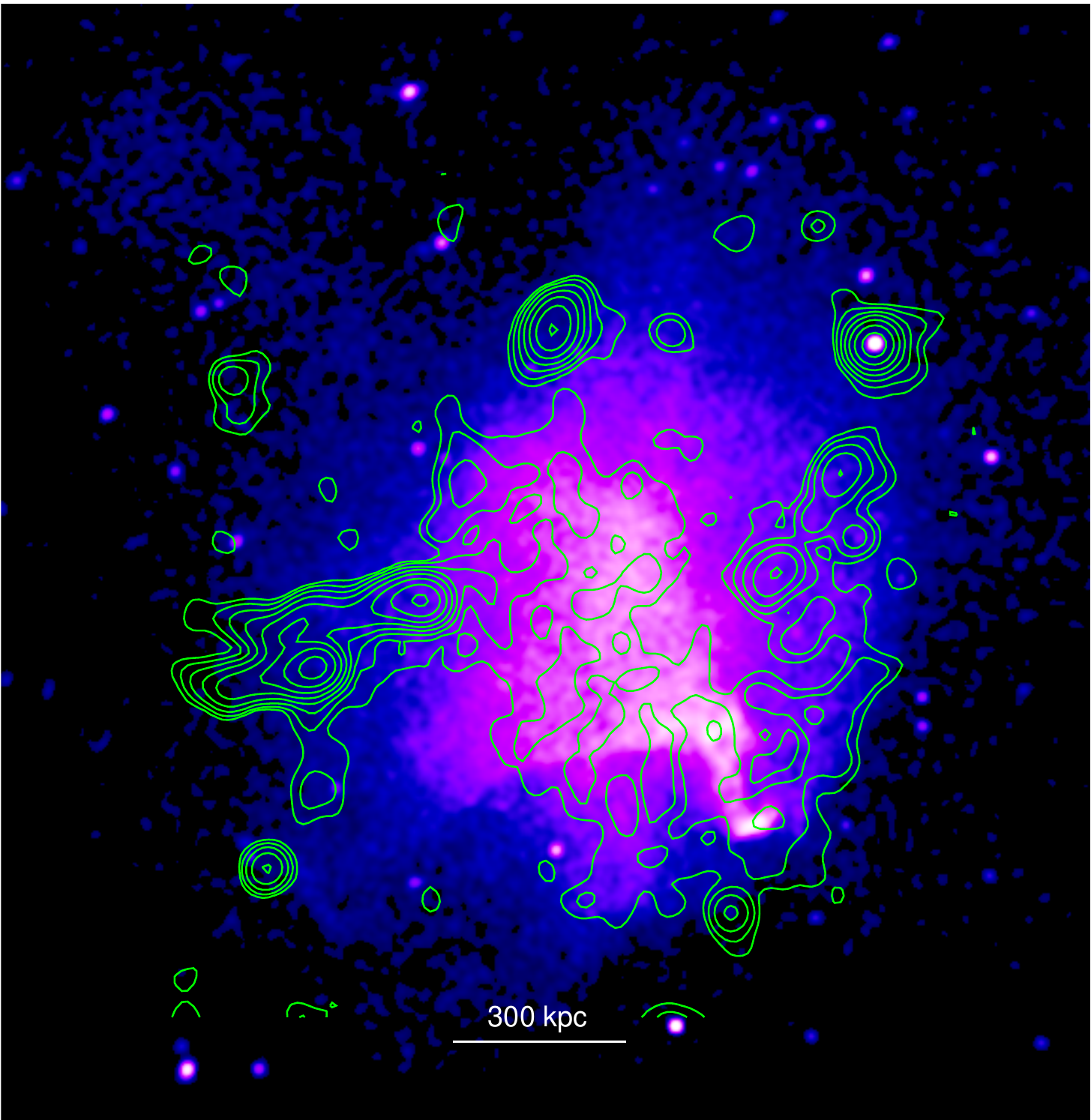}
 \figsubcap{A520}
 }
 \vspace{1.5mm}

 \parbox{2.1in}{%
  \includegraphics[height=5.2cm,bb=24 14 277 267,clip]%
    {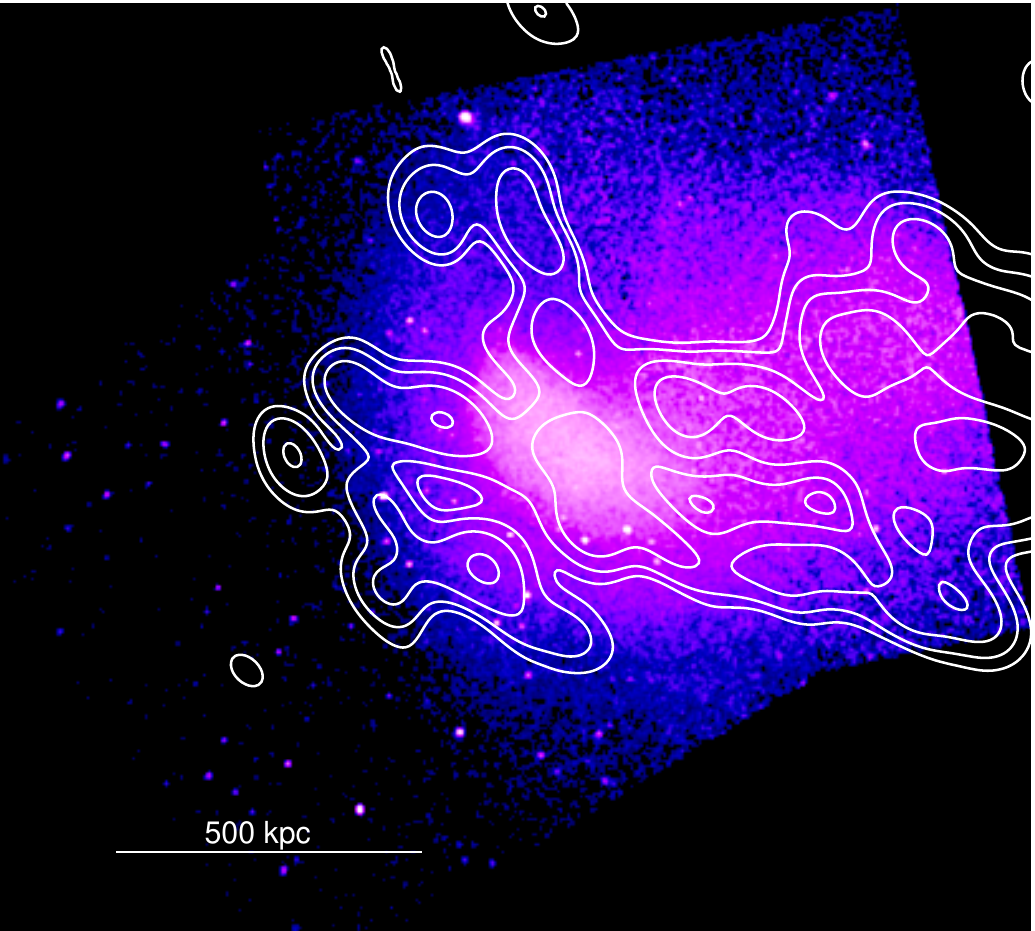}
 \figsubcap{A754}
 }
 \hspace*{4pt}
 \parbox{2.1in}{%
  \includegraphics[height=5.2cm,bb=1 1 270 270,clip]%
   {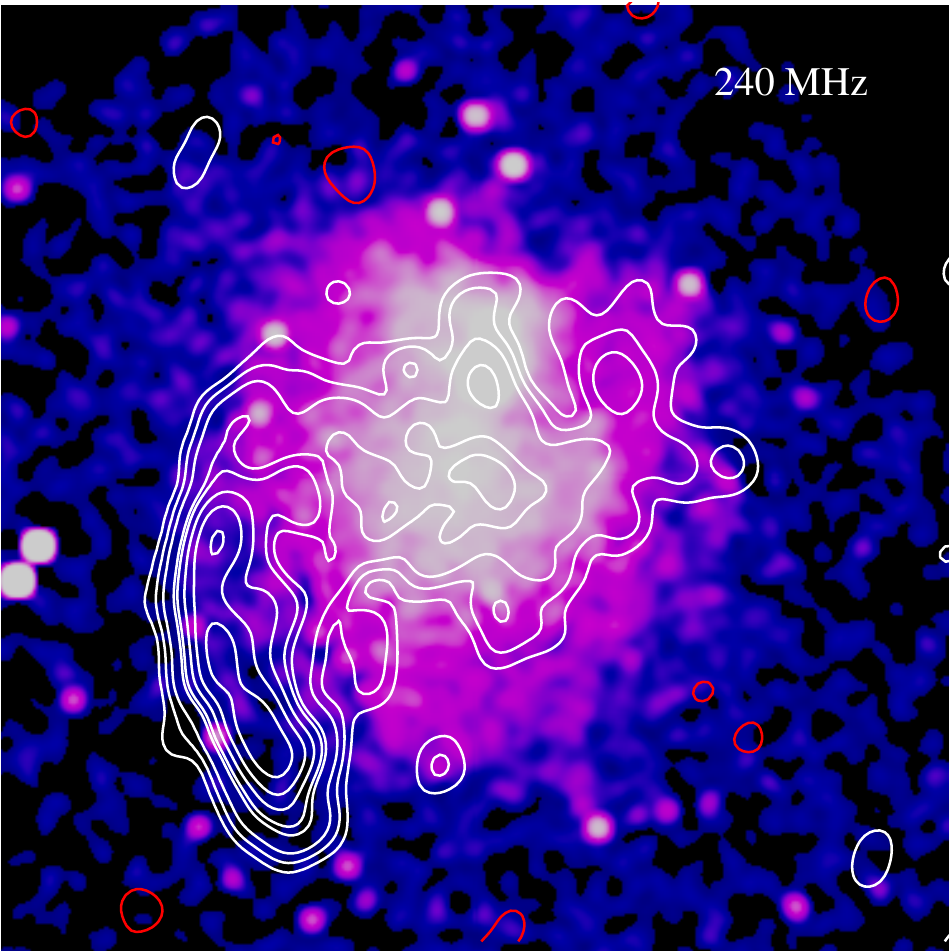}
 \figsubcap{A521}
 }
 \vspace{1.5mm}

 \parbox{2.1in}{%
  \includegraphics[height=5.2cm,bb=110 90 520 500,clip]%
    {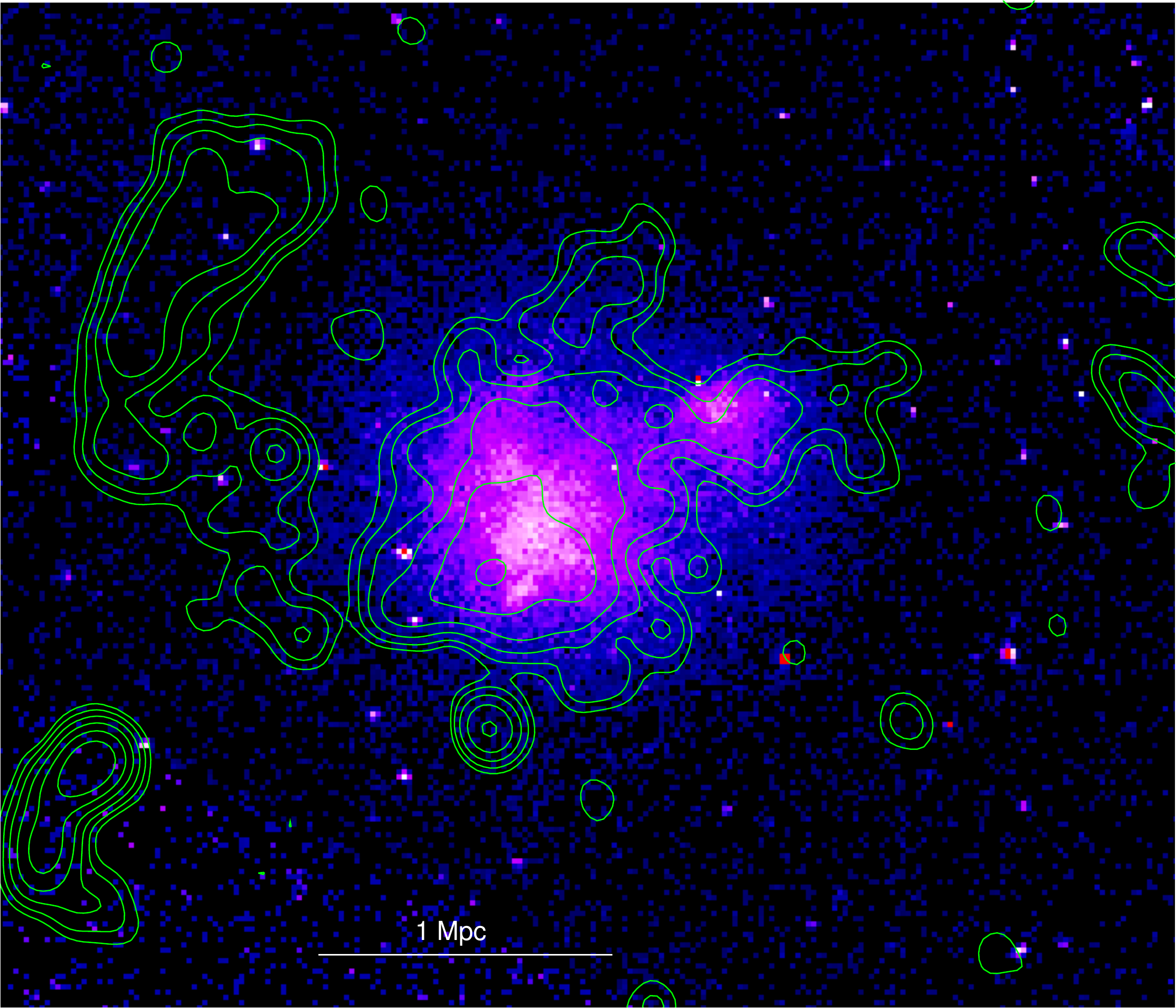}
 \figsubcap{A2744}
 }
 \hspace*{4pt}
 \parbox{2.1in}{%
   \includegraphics[height=5.2cm,bb=68 78 418 428,clip]%
    {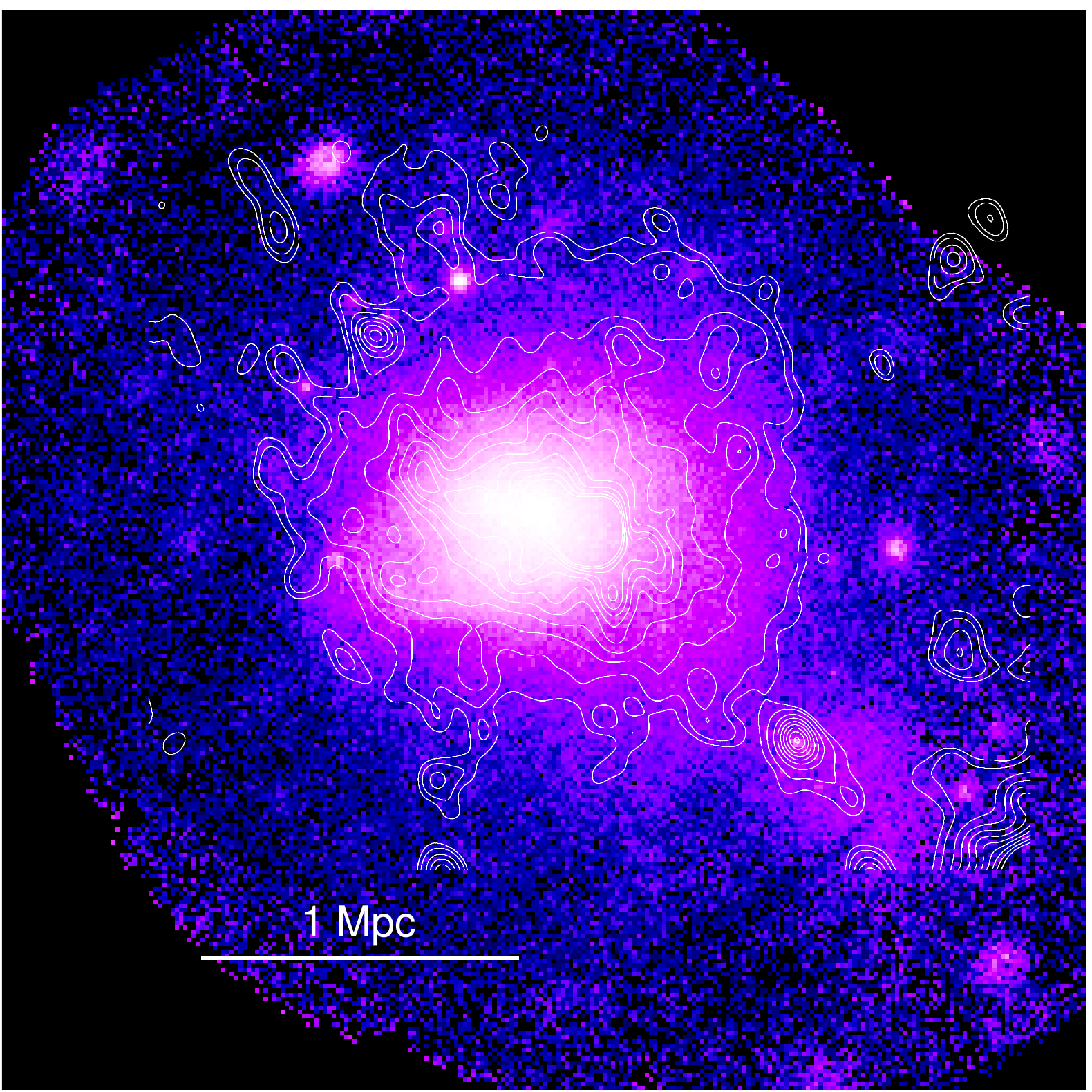}
 \figsubcap{Coma}
 }

 \caption{Contours of radio halos overlaid on X-ray images of clusters with
   shock fronts (\1e\cite{liang00,mm02}, A520\cite{govoni01,mm05},
   A754\cite{macario10}) or front candidates (A521\cite{brunetti08},
   A2744\cite{venturi10}, Coma\cite{brownrudnick10}).  In all cases,
   there is coincidence of a shock front seen in the X-ray with an edge in
   the radio halo.}
\label{fig:rhalos}
\end{center}
\end{figure}

Merger shock fronts are expected and, as we saw above, observed to have
relatively low Mach numbers. Indeed, a test particle falling into the
gravitational potential of a typical cluster would acquire $M\sim 3-4$.  An
infalling subcluster would generate multiple shocks in front of it,
preheating the ambient gas and further reducing its Mach number by the time
it arrives into the cluster's X-ray bright central region, where we observe
it.  Such weak shocks are believed to be inefficient accelerators of
electrons from the thermal pool to Lorentz factors $\sim 10^4$ required to
produce relics and radio halos\cite{kang02,macario10}.  Yet, it has been
noticed that synchrotron radio halos in \1e\cite{mm02} and A520\cite{mm05}
display brightness edges coincident with the X-ray shocks (Fig.\
\ref{fig:rhalos}), indicating that these shocks must have something to do
with producing radio emission at least at those locations --- either
accelerating electrons, or compressing or re-accelerating a pre-existing
relativistic electron population, and perhaps strengthening the magnetic
fields beyond simple compression\cite{mm05}.  Re-acceleration of
pre-existing relativistic electrons is an attractive theoretical
possibility\cite{brunetti01,kang05}.

All the clusters with newly discovered X-ray shocks and shock candidates
exhibit radio halos or relics, except A2146 that doesn't have sensitive
radio data yet. It is interesting to overlay the radio images on the X-ray
shocks, which we do in Fig.\ \ref{fig:rhalos}. In addition to \1e\ and A520
discussed above, it shows A521, which exhibits a relic coincident with the
probable shock, plus a halo extending from that relic across the whole
cluster\cite{giacintucci08,brunetti08} --- a radio morphology very similar
to A520. A754, with the weakest shock ($M=1.6$), exhibits a steep-spectrum
relic at 74 MHz\cite{kassim01}, coincident with the shock front, and a radio
halo extending from the relic westward across the
cluster\cite{bacchi03,macario10}.  Finally, the two shock candidates in
A2744 and Coma, which we report in this paper, both have edge-like features
seen in recent high-quality radio images of their
halos\cite{venturi10,brownrudnick10}, coincident with putative X-ray shock
fronts. In Coma, the X-ray edge spans the southern half of the radio edge
(Fig.\ \ref{fig:rhalos}).  The cluster RXJ\,1314--25 (not shown) also has a
relic at the shock front and and a small radio halo extending from
it\cite{mazzotta10}.

Thus, spatial coincidence of the merger shocks with edges of radio halos ---
or with radio relics that delineate an edge of the radio halo --- is quite
ubiquitous. It supports the scheme that we proposed for A520 and
A521\cite{brunetti08}, in which a weak merger shock re-accelerates
pre-existing relativistic electrons (which may remain from earlier mergers,
or generated by collisions of the long-lived cosmic ray protons with thermal
protons\cite{miniati01}).  As these electrons move downstream from the
shock, they rapidly cool, creating a narrow arc-like ``relic'', but at a
certain distance are picked up and re-accelerated again by turbulence
generated by the merger behind the shock, which produces the cluster-wide
halo. In this picture, the edge of the radio halo and the bulk of the halo
are distinct phenomena, though both caused by the same merger.  Depending on
the Mach number of the shock, the magnetic field behind the shock (that
determines the rate of synchrotron cooling of the electrons and thus the
width of the ``relic'' in the direction across the shock front), and the
power spectrum of turbulence in the body of the ICM behind the shock, the
relic may appear at some frequencies as a distinct radio source, while at
other frequencies, merge with the halo. In A521, the relic is dominant at
$\nu>1$ GHz, while in A754, the relic dominates at 74 MHz but merges with
the halo at higher radio frequencies.  The prediction of this picture is
that the radio spectrum at the halo ``edge'' should be a power law
determined by the shock's Mach number (Fermi acceleration from thermal pool
and re-acceleration from plausible initial electron distributions result in
the same spectrum). This is again consistent with observations of
A521\cite{giacintucci08} and A754\cite{macario10}.  The radio spectrum of
the rest of the halo (more exactly, the frequency of its exponential cutoff)
is determined by the velocity of the turbulence and the strength of the
magnetic field\cite{brunetti07}. Depending on that spectrum, it may or may
not be possible to see the steepening of the radio spectrum of the front edge
due to aging of electrons as one moves from the shock front inward (as was
possible for A521\cite{giacintucci08}).

We note that some more irregularly shaped cluster radio relics, such as a
bright relic in A2744 (seen in the left edge of the image in Fig.\ 
\ref{fig:rhalos}), A2256 or A3667, may have different origin, perhaps
involving a shock passage across a distinct region of fossil radio
plasma\cite{ensslin01}.

It is clear that detailed, spatially resolved studies of cluster merger
shocks in the X-ray and at multiple radio frequencies is a promising way to
study the cosmic ray acceleration mechanisms in astrophysical plasmas.  We
may be starting to collect a sufficiently large sample for such studies.



\end{document}